# Spectroscopic characterization of magnetoplasmonic nanodisk array: size, shape and lattice constant

G. X. Du*, S. Saito, and M. Takahashi

Department of Electronic Engineering, Graduate School of Engineering, Tohoku University, 6-6-05 Aoba, Aramaki, Aoba-ku, Sendai 980-8579, Japan

Email: guanxiang.du@unibas.ch

**Abstract.** Magneto optical activity implemented in plasmonic nanostructures provides a new functionality in active plasmonics. To what an extent, one can benefit from this new degree of freedom is still under open discussion. In this work, with the development of a wavelength parallel characterization tool for measuring the optic and MO properties of nanostructures, we present systematic studies of the magneto-optical activity of Au/[Co/Pt]$_n$/Au nanodisk array with various of disk size, shape and grating constant.

## Introduction

With the rapid growth of plasmonics [1], implementing magneto optical (MO) functionality into plasmonic nanostructure has brought forth the emerging of a sub-field, now termed as magnetoplasmonics [2 – 5, 5a]. Researches has found a number of interesting functionality between the interplay of plasmonic and MO activity. Examples include enhanced MO effect associated with plasmon resonance [6], manipulation of plasmon modes via external magnetic field in MO active plasmonic devices [7]. Most of the current magnetoplasmonic materials rely on the use of ferromagnetic metals to obtain a sizable MO activity. Despite its high activity to external magnetic field and fast response associated with the time scale of magnetization reversal (nanosecond to picosecond easily accessible), the main drawback of the current 3d transition ferromagnetic metals is their highly dissipative nature that suppresses plasmonic strength. This is currently a big stone on the way towards potential applications [8, 8a] and several studies have addressed this issue [8b,8c]. With rich resonance features in magnetoplasmonic materials, the MO spectra in the visible range covering plasmon resonances are highly desirable. Most groups, including the earlier studies in our group, rely on the use of a modulation method to measure MO properties, where a monochromator is used to get wavelength resolution. [2 - 9] Since this method provides only MO properties at a single wavelength in every magnetic field scan, it is thus quite a time consuming work to study the spectroscopic phenomena in magnetoplasmonic materials. This is even heavier task when requiring resolving very narrow resonance lines where one needs a large number of wavelength-scan. Motivated by the desire for a fast acquisition of MO spectra, we have developed a fast MO spectrometer using white light source and spectrometer as the detector [10]. The measurement system is surprisingly simple, low cost and easy to make without the need for light polarization modulation. In our previous work, the dependence of size [17] and shape [22] of nanodisks on the magnetoplasmonic properties was studied but limited to several laser wavelengths. In this focus, we present the recent results on the spectroscopic characterization of optical and MO properties of magnetoplasmonic Au/[Co/Pt]$_n$/Au nanodisk array with various disk size, shape and grating constants. The measurement was done in a rather time efficient way. As a reference, plasmonic nanodisk array using only gold is also shown to reveal general plasmon features, its size and grating dependence, with much sharper resonance as compared to the much broadened features in sandwich nanodisk array. A direct comparison between them give a clear illustration of how dissipative nature of ferromagnetic metal suppresses the plasmonic resonance strength.

**Measurement setup and layout of nanodisk array**

We developed a microspectrometer to characterize the spectroscopic optical and magneto optical properties of nanostructures in a wavelength-parallel way. It works in both reflection and transmission configuration. Fig. 1 (a) is a schematic view of the setup in transmission case. The details of the setup can be found in ref. 10. Briefly, it uses white light source and a spectrometer, with additional polarizing optical components. Lenses allow white light from fiber to be focused to micrometer-scaled spot that is necessary for characterizing nanostructure ensembles. Wavelength-parallel measurement enables fast acquisition of spectroscopic data. Fig. 1 (b) shows the nanodisk array studied in this work, which is fabricated by electron beam lithography (EBL) and ion milling. Nanometer sized disks (diameter, $d$) are arranged in a square lattice. The nearest neighboring disks are spaced by $h$, the lattice constant. A photoresist with thickness of 150 nm is coated as a cap layer, refractive index of which is matched to the glass substrate, creating a homogeneous dielectric environment for nanodisks.

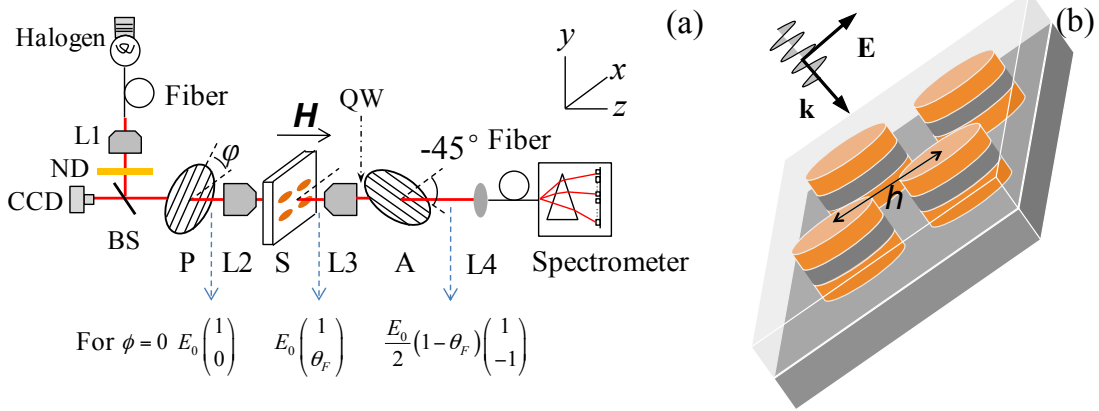

Fig. 1 (a) Schematic view of the setup. Jones vectors of light after the polarizer (P), sample (S) and analyzer (A) are shown, with transmission axis of polarizer in the horizontal plane ($\varphi = 0$). (b) Illustration of the nanodisk array in a square lattice with lattice constant of $h$. nanodisks sit on glass substrate and covered by photoresist thin film that forms a homogeneous dielectric surroundings for nanodisks.

In this work, nanodisks with stacking structure of Au and Au/[Co/Pt]$_n$/Au sandwich are studied. Gold nanodisks, compared to Au/[Co/Pt]$_n$/Au sandwich, exhibiting sharper plasmonic resonance, are investigated to show clear plasmon resonance in both reflection and transmission configuration and a narrower plasmon resonance with increasing lattice constant. The stacking structure for gold and sandwich nanodisks are Ti,2/Au,40 and Ti,2/Au,20/[Pt,1/Co,0.5]$_7$/Au,20 (nm), respectively. The Faraday spectra of the sandwich continuous thin film can be found in Fig. 1S.

**Optical properties of gold nanodisk array**

Top-down nanofabrication technique based on EBL and ion milling provides us a wide tunability on plasmon modes by varying the nanostructure size, arrangement and shape. Researchers can certainly benefit from this aspect in the interplay between rich plasmon modes and MO activity. Examples are, extradinary optical transmission [12] in periodically arranged subwavelength holes in an otherwise opaque metallic thin film, multipolar Fano resonance [13] and electromagnetic induced transparency [14]. In this work, the electromagnetic coupling between nanodisks in the array is directly observable when varying the lattice constant.

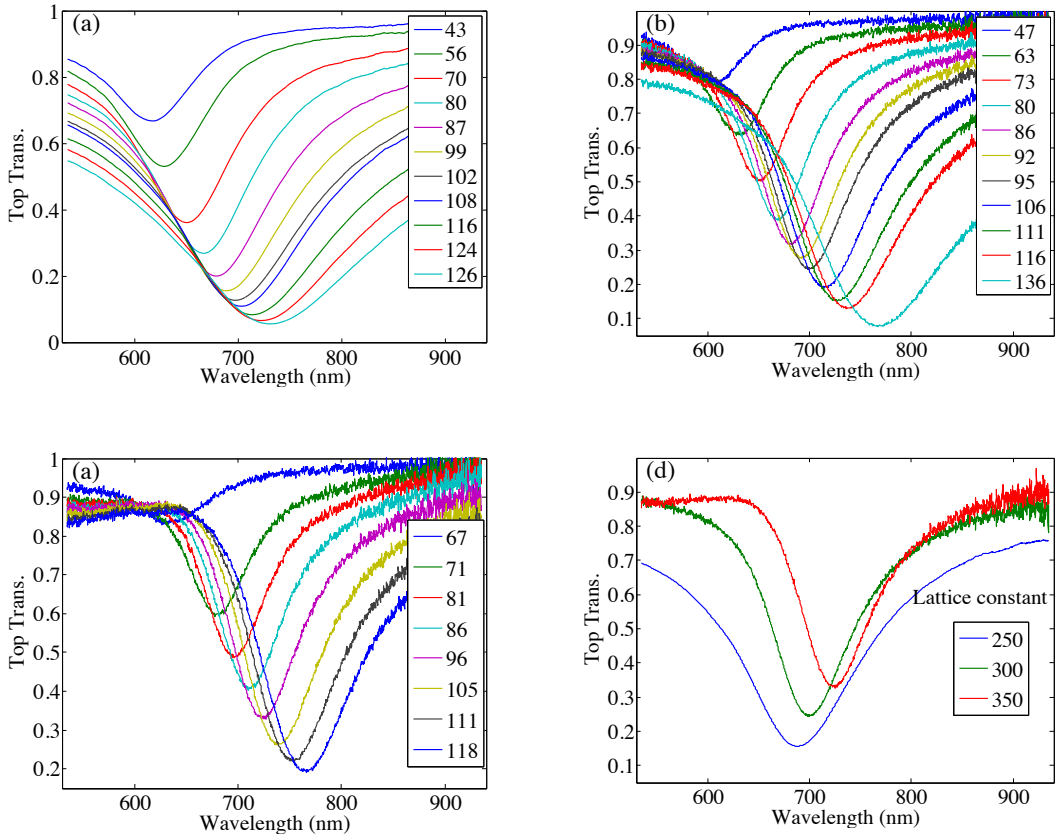

Fig. 2 Narrowing of plasmon resonance with longer lattice constant $h$. Top transmittance for gold nanodisk array with $h$ = 250 (a), 300 (b) and 350 (c), unit in nm, each with various disk diameters. (d) Direct comparison of the effect of lattice constant on the resonance line width, with nearly the same disk size $d$ = 97 ± 2 nm.

Fig.2 (a - c) show the transmittance spectra of gold circular nanodisks of various diameters, with lattice constant $h$ = 250, 300, 350 nm, respectively. Light is incident from the nanodisk side (top incidence). Reference for transmittance measurement uses the light passing through a bare glass substrate. The dips in transmittance spectra indicate plasmonic resonance, which is redshifted with increasing nanodisk size, known as retardation effect [11] and is associated with the finite size of disks compared to the excitation light wavelength. A general feature is the narrowing of the resonance linewidth as increasing lattice constant. This narrowing effect is found to be even sharper when extending the lattice constant to 400 – 600 nm [15]. At this long lattice constant region, the plasmon resonance approaches and interferes with the Rayleigh anomaly (first order is at $\lambda_0 = nh$, $n$, the refractive index of dielectric surroundings, $h$, the lattice constant). As a direct comparison, Fig. 2 (d) shows the transmittance spectra of the three lattice constants. The disk diameters are chosen to be almost the same though slightly different, 99, 96 and 95 nm, correspondingly. Nanodisk array with lattice constant of 350 nm shows a much narrower resonance dip compared to that with a lattice constant of 250 nm. As pointed out in Ref. 15, this narrowing feature holds promise in ultrasensitive refractive index sensing. The figure of merit (peak position shift per unit change in refractive index divided by the resonance linewidth) is much higher with significantly narrow resonance line.

**Optical and MO properties of Au/[Co/Pt]$_n$/Au circular nanodisk array**

Plasmon enhanced magneto optical effect is clearly seen in both reflection (Kerr) and transmission (Faraday) configuration, as shown in Fig. 4 (a - c). Nanodisk array consists of nanodisks arranged in a square lattice with lattice constant of 250 nm, with a stacking structure of Au/[Co/Pt]$_n$/Au. Plasmon resonance appears as dips in the transmittance spectra, though not as sharp as the resonance features in gold nanodisk array shown in Fig. 3 (a). Peak positions of ellipticity are correlated with those of plasmon resonances. This correlation has been observed in our previous work [17] at several wavelengths of laser diodes. Here we provide further characterization in the visible to near infrared spectra. At the wavelength where Faraday ellipticity takes maxima, the rotation angle cross zero, which is characteristic of resonance, usually termed as Kramers–Kronig relations. [18] With increasing disk size, the amplitudes of Faraday rotation and ellipticity increase, associated with a loss in transmittance. This is also of typical resonance feature and found in many other works [19, 20]. Plasmon enhanced Kerr effect was also observed. Due to the highly dispersive nature of [Co/Pt]$_n$ multilayers, tips in the reflectance curves are much broadened compared to gold nanodisks, however, resonant MO is still quite clear in Fig. 4 (e – f), especially for disk size of 148 nm. To be noted, the significant large Kerr effect seen in disk size of 84 nm reaching 1 degree at around 700 nm, is a combined contribution from Kerr rotation and Faraday rotation, with the latter doubled due to back reflection at the glass/air interface (4%). The electromagnetic coupling between nanodisks in this example is out of the scope of this work. Briefly addressing, Bragg diffraction which is expected to be important at the first diffraction order $\lambda_0$ = 375 nm is not observable in the wavelength range from 530 to 940 nm.

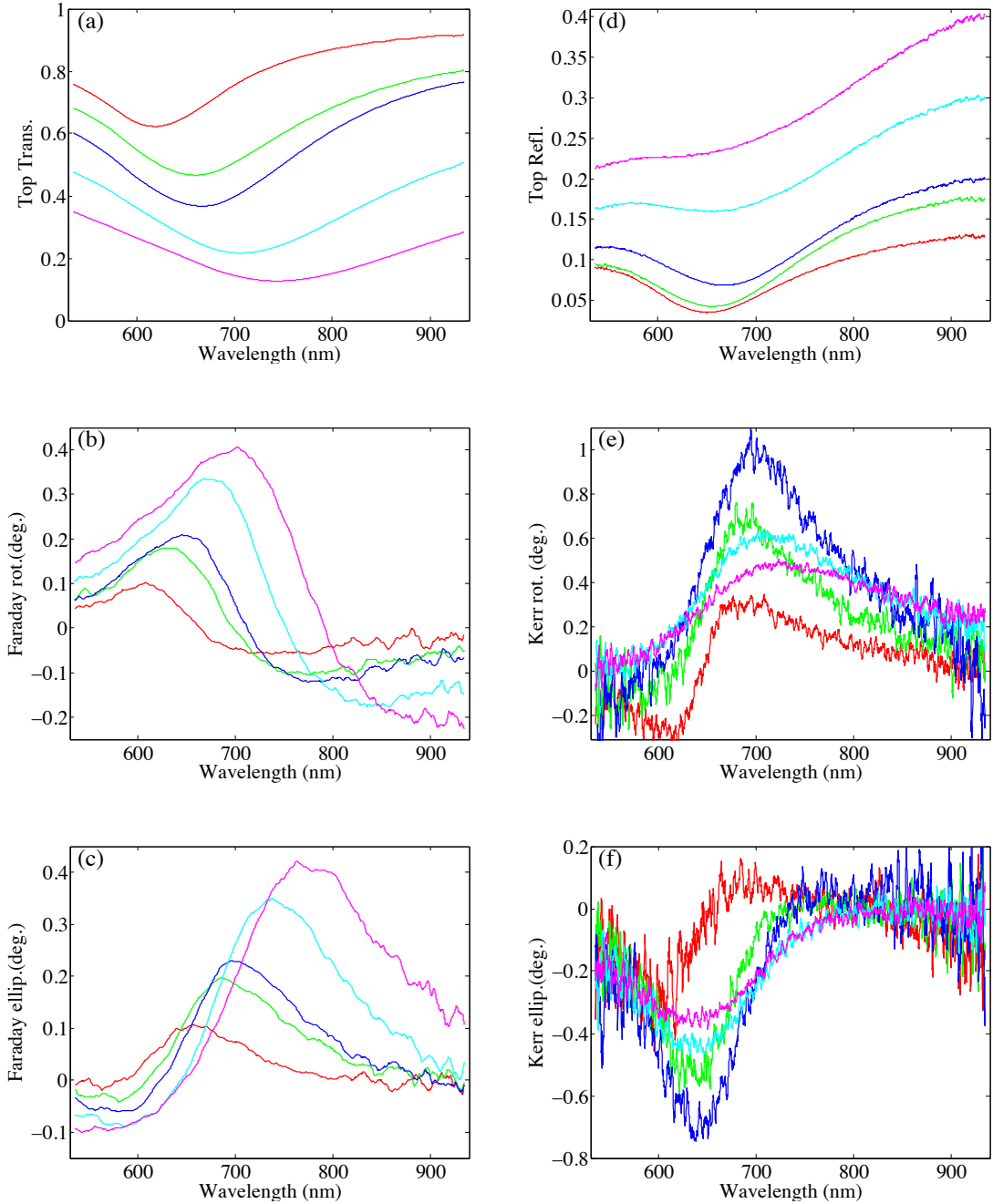

Fig. 4 Faraday (a – c) and Kerr activity (d – f) of Au/[Co/Pt]$_7$/Au nanodisk array. Each peak in Faraday ellipticity in (c) is correlated with the plasmon resonance in (a). Reflectance spectra display much broadened resonance features. Comparing with the transmittance spectra reveals the back reflection at the glass/air interface, which partially contributes the total MO effect in Kerr configuration with light pass the nanodisks twice. Surprisingly large Kerr rotation for $d$ = 84 nm at 700 nm (e) is therefore a combined Kerr and Faraday effect. Data for Faraday effect was smoothed. Each line color corresponds to the same disk size.

This back reflection exhibits a dip in the top reflectance spectra, which is otherwise absent at bottom incidence (light passes through glass substrate then nanodisks). The dip position in the reflectance and transmittance spectra is correlated, redshifting with increasing disk size, The contribution to the total reflectance from black reflected light follows the expression $0.04T^2$, where the coefficient 0.04 comes from the reflectance at glass/air interface and $T^2$ (square of transmittance) accounts for the double pass of light through nanodisks. This back reflection makes significant contribution to the total reflected light especially when the front reflection from nanodisks is smaller than 10%. A clear difference in the reflectance spectra for top and bottom incidence is shown in Fig. 5. Data is taken with gold nanodisks. The dip in top transmittance and tip in bottom reflectance are correlated to plasmon resonance. The dip in top reflectance spectra is observed that is absent in the bottom reflectance.

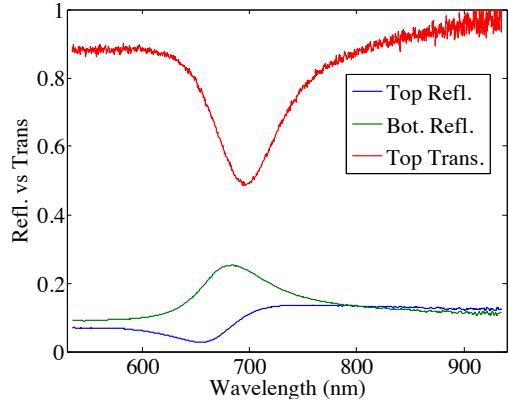

Fig. 5 Evidence of back reflection at glass/air interface: a dip in the top reflectance spectra at 660 nm. Data is from gold nanodisk array with $(d, h)$ = (81, 350) nm.

**Optical and MO properties of Au/[Co/Pt]$_n$/Au elliptical nanodisk array**

Shape anisotropy provides a degree of freedom to tune the plasmon resonance, giving rise to polarization dependent optical response and has been used in ultrasensitive refractive index sensing. [21] Fig. 6 shows the optical and Kerr effect of Au/[Co/Pt]$_n$/Au elliptical nanodisk array ($h$ = 250 nm), the long (short) axis for the right and left panels are 96 (65) and 156 (73) (unit: nm), respectively. Measurements with light polarization along the two axes were performed. This polarization dependent MO effect is quite different from conventional MO thin film which has isotropic optical refractive index in the film plane thus displaying isotropic MO effect, while in the anisotropic nanodisks, optical response is polarization dependent, giving rise to the observed intriguing anisotropic MO effect. The large difference in both optical and MO properties between the two orthogonal polarizations is associated with different plasmon modes, i.e., transverse and longitudinal modes, which is much less pronounced in anisotropic dielectric nanostructures (experimentally verified, data not shown). The interplay between optical anisotropy and MO effect in anisotropic nanodisk has been formulated in Ref. [22], while here we present its spectroscopic characterization. To be noted, in contrast to circular nanodisks, one could not simply correlate the dips or tips in the Kerr spectra with that of the reflectance spectra.

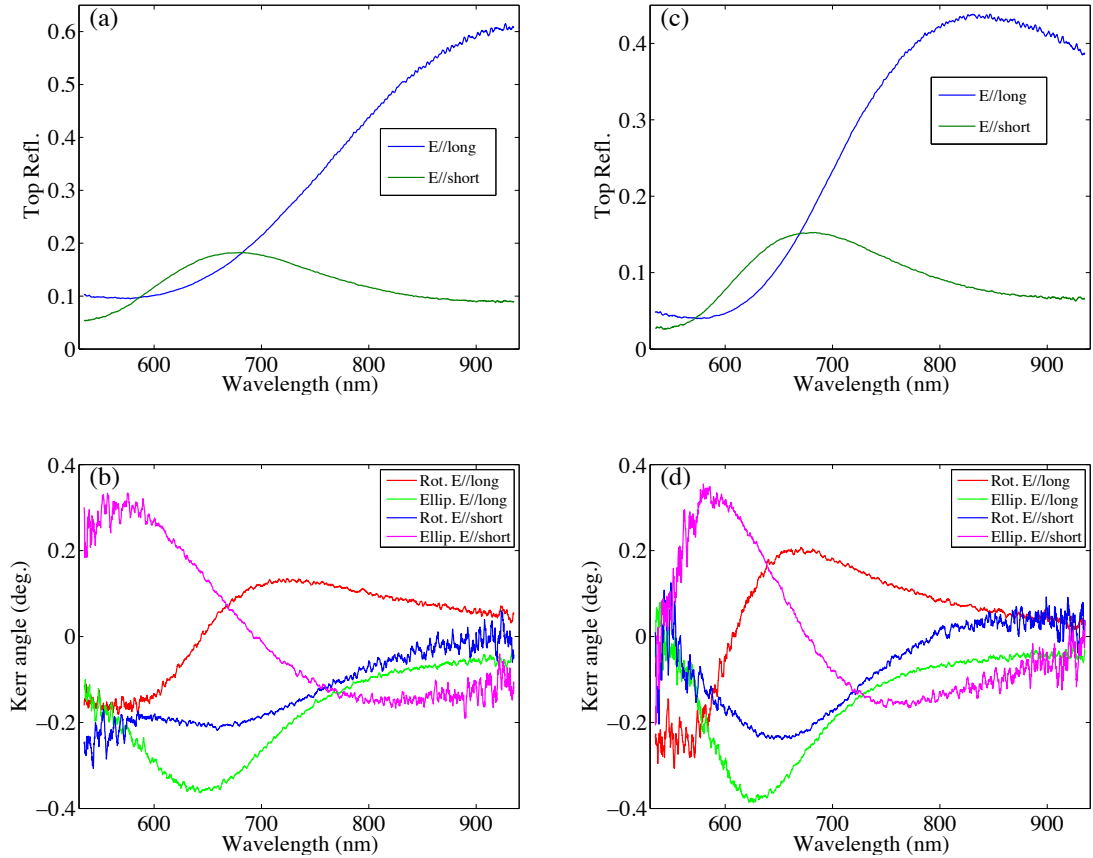

Fig. 6 Anisotropic optical reflectance and Kerr activity of Au/[Co/Pt]$_n$/Au elliptical nanodisk array. The long (short) axis lengths are 156 (73) and 96 (65) nanometer for elliptical nanodisk in left (a – b) and right (c – d) panel, respectively. Measurements were performed with light polarization (E) along the long or short axis.

**Conclusion and outlook**

In this work, we present a spectroscopic characterization of Au/[Co/Pt]$_n$/Au nanodisk array. A systematic investigation on the size and shape dependence was performed. Resonance features were observed in circular nanodisk array, in both Kerr and Faraday spectra. The positions of the resonances were correlated with the localized surface plasmon resonance. Polarization dependent Kerr spectra in elliptical nanodisk array were presented. In the first part of this work, we also showed the optical spectra of gold nanodisk array, where we figure out that the electromagnetic coupling between nanodisks could results in a narrowed plasmon resonance. This narrowing effect will be addressed in the further studies by implementing MO activity. Back reflection at the substrate/air interface was revealed as an important technical issue in the characterization of nanodisk/dielectric-substrate system, especially in the top reflection, in the small reflectance (<10%) region.

**Acknowledgement**

Support from the foreign postdoctoral program from JSPS (Grant-in-Aid No. 2109282) is acknowledged.

**Appendix**

Fig. 1S shows the Faraday rotation and ellipticity angle of the sandwich thin film.

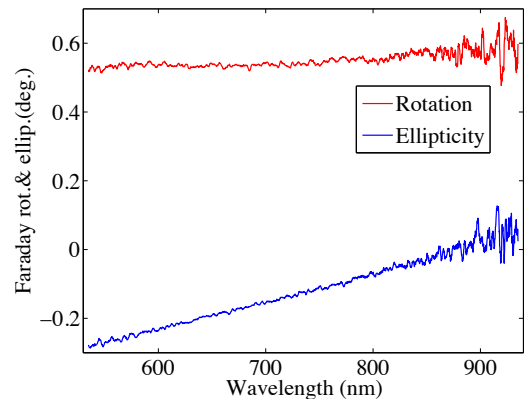

Fig. 1S Faraday rotation and ellipticity of Au/[Co/Pt]$_n$/Au thin film.